\documentclass{article}

\usepackage[final]{neurips_2021}

\usepackage[utf8]{inputenc} 
\usepackage[T1]{fontenc}    
\usepackage{hyperref}       
\usepackage{url}            
\usepackage{natbib}
\usepackage{booktabs}       
\usepackage{amsfonts}       
\usepackage{nicefrac}       
\usepackage{microtype}      
\usepackage{xcolor}         
\usepackage{enumitem}
\usepackage{amssymb}
\usepackage{pifont}
\usepackage{diagbox}
\usepackage{multiaudience}

\usepackage{amssymb}
\usepackage[small]{caption}
\usepackage{graphicx}
\usepackage{amsmath}
\usepackage{amsthm}
\usepackage{booktabs}
\usepackage{algorithm}
\usepackage{algorithmic}
\usepackage{amsmath} 
\usepackage{subfigure}
\usepackage{multirow}
\usepackage{booktabs}
\usepackage{wrapfig}

\usepackage{footmisc}

\title{MemFine: Memory-Aware Fine-Grained Scheduling for MoE Training}

\author{
ZTE AIH Team, \quad INESA-ZTE Joint Laboratory
}

\begin{document}

\noindent  

\maketitle

\begin{abstract}
The training of large-scale Mixture of Experts (MoE) models faces a critical memory bottleneck due to severe load imbalance caused by dynamic token routing. This imbalance leads to memory overflow on GPUs with limited capacity, constraining model scalability. Existing load-balancing methods, which cap expert capacity, compromise model accuracy and fail on memory-constrained hardware. To address this, we propose MemFine, a memory-aware fine-grained scheduling framework for MoE training. MemFine decomposes the token distribution and expert computation into manageable chunks and employs a chunked recomputation strategy, dynamically optimized through a theoretical memory model to balance memory efficiency and throughput. Experiments demonstrate that MemFine reduces activation memory by $48.03\%$ and improves throughput by $4.42\%$ compared to full recomputation-based baselines, enabling stable large-scale MoE training on memory-limited GPUs.
\end{abstract}
\section{Introduction}
With the exponential growth of artificial intelligence model sizes, traditional dense models face severe challenges in computation and memory~\cite{wang2025step}. The MoE model, as a sparse architecture, introduces a gating network and multiple expert sub-networks, significantly reducing the activated computation per forward pass while maintaining a massive number of model parameters~\cite{shazeer2017outrageously}. This characteristic makes it possible to build and train giant models with trillions or even tens of trillions of parameters under limited computational resources, rapidly becoming a key technical path driving the development of large language models and tasks in other domains~\cite{zhang2022data,yin2022deep,zhang2023dual}. Therefore, achieving efficient and stable large-scale MoE training has become an urgent demand in both industry and academia~\cite{yang2025qwen3, rajbhandari2022deepspeed}.

Although MoE models possess significant theoretical advantages, they face the challenge of load imbalance during practical training, especially in large-scale distributed environments~\cite{cai2025survey}. To maximize the expressive power and performance of the models, modern large-scale MoE training allows the number of tokens processed by each expert to be unrestricted. However, this strategy, combined with the stochastic routing of the gating network, leads to a highly non-uniform distribution of tokens among different experts~\cite{lepikhin2020gshard, wang2024auxiliary}. This load imbalance phenomenon further translates into severe imbalance in GPU memory consumption. 
Specifically, during training, some GPUs might be assigned a number of tokens far exceeding their memory capacity, causing them to encounter memory overflow first and thus interrupting the entire training job. This issue severely constrains the potential for scaling MoE model size on GPU clusters with limited memory capacity~\cite{kim2024scaling}.

To address the aforementioned memory bottleneck, existing research primarily employs load-balancing solutions by setting a capacity factor to limit the number of tokens processed by each expert to a balanced level~\cite{lepikhin2020gshard, wang2024auxiliary,fedus2022switch,liu2024deepseek}, while applying activation recomputation~\cite{chen2016training,korthikanti2023reducing} and offloading strategies~\cite{ren2021zero}. Although this method can effectively avoid memory peaks, it severely disrupts the inherent dynamic routing mechanism of the MoE model, leading to degraded training accuracy and failing to realize the full potential of the MoE architecture. More importantly, these methods are generally applied on GPUs with large memory capacity and cannot effectively tackle the memory overflow problem caused by extreme load imbalance on GPUs with small memory capacity.

To overcome the above challenges, this paper proposes a memory-aware fine-grained MoE scheduling framework, named MemFine. MemFine firstly designs a fine-grained token dispatch-expert computation-token combine algorithm, which divides the token distribution process into multiple chunks and introduces a chunked recomputation optimization during the backward pass to maximize memory savings. Furthermore, to minimize the impact of chunking on end-to-end training performance, MemFine designs an adaptive chunk tuning algorithm based on a theoretical memory cost model, achieving dynamic chunk splitting according to the training progress and maximizing the effective end-to-end computational throughput. MemFine enables trainers to successfully tame the memory peaks under large Expert Parallelism (EP) training without touching the routing logic, thereby empowering the training of ultra-large-scale MoE models on GPUs with small memory capacity. The innovations of this paper are summarized as follows:

\begin{enumerate}[label=\textbullet]
    \item This paper proposes a fine-grained scheduling framework, MemFine, to solve the Out-Of-Memory (OOM) problem caused by uneven token distribution in MoE models. MemFine effectively reduces the memory demand of model activations through the fine-grained token distribution and recomputation, enabling the training of MoE models with large EP on GPUs with small memory capacity.
    \item This paper introduces a memory-aware dynamic tuning algorithm within MemFine, which balances memory and performance by controlling the number of chunks for token distribution through threshold adjustment.
    \item Experiments validate the effectiveness of the design of the proposed MemFine. Compared to the traditional MoE distribution method combined with full activation recomputation, MemFine achieves a $48.03\%$ reduction in activation memory usage while increasing throughput by $4.42\%$.
\end{enumerate}

\section{Related Work}
\subsection{MoE Training and Expert Parallelism} 
The MoE model has emerged as a key technology in recent years to address the scalability challenges of ultra-large-scale models~\cite{cai2025survey}. Its core idea involves replacing the traditional dense feedforward network layer with a structure composed of multiple experts, i.e., small feedforward networks, and introducing a gating network that dynamically selects and activates a subset of these experts for each input token. This conditional computation paradigm enables a dramatic increase in the total number of model parameters while keeping the actual computational cost per token constant, thereby achieving an effective trade-off between computational efficiency and model capacity. Representative works~\cite{yang2025qwen3, lepikhin2020gshard, wang2024auxiliary, fedus2022switch, liu2024deepseek} have successfully scaled the model parameter count to the trillion-level.

To efficiently train and deploy MoE models, EP has been developed as a specialized model parallelism strategy~\cite{jin2025megascale}. Unlike traditional model parallelism, which splits a single model layer across different devices, EP distributes different experts to distinct computing devices. The routing decisions of the gating network determine the flow of input data across devices, with each device only computing the experts it hosts. This paradigm has been adopted and optimized by mainstream open-source frameworks~\cite{rasley2020deepspeed,shoeybi2019megatron,zheng2024llamafactory}. 


\subsection{Load Unbalance}
In MoE models, expert load imbalance is a core challenge. If the gating network consistently routes a large number of inputs to a few hot experts while other experts remain under-utilized and under-trained, it leads to wasted model capacity, training instability, and ultimately performance degradation. To address this issue, researchers have proposed various methods combining the soft and hard constraints.

Current mainstream approaches primarily involve introducing an auxiliary load balancing loss into the loss function to regularize the gating network. The core idea of these methods is to encourage the average activation frequency of all experts to become more uniform. A classic early work is the auxiliary load balancing loss adopted in Switch Transformer~\cite{fedus2022switch}. It calculates the within-batch average routing probability for each expert and compares it with the average probability across all experts, using a mean square error loss to penalize deviations, thereby smoothly steering routing towards less loaded experts. To mitigate the accuracy impact of the auxiliary loss, DeepSeek~\cite{wang2024auxiliary,liu2024deepseek} proposed an auxiliary loss-free load balancing algorithm, which dynamically adjusts the routing bias of experts to control token distribution. While these methods can effectively alleviate the overall load imbalance phenomenon, they still cannot avoid a small number of iterations with extreme load imbalance.

Another important approach combines hard constraints with soft incentives. For example, GShard~\cite{lepikhin2020gshard}, alongside an auxiliary loss, introduced the concept of expert capacity, setting an upper limit on the number of tokens each expert can process. When an expert reaches its capacity limit, excess tokens are forcibly routed to less loaded experts. This serves as an effective hard guarantee mechanism to prevent individual expert overload, but this method has been verified to affect the convergence of the overall training loss~\cite{tang2025pangu}.

Furthermore, activation memory optimization techniques such as full recomputation and activation offloading are often combined with the aforementioned methods to save memory and reduce the impact of distribution imbalance~\cite{liu2024deepseek}.

\begin{table}[t]
\caption{Notation table}
\label{notationlist}
\vspace{5pt}
\scalebox{0.8}{
\begin{tabular}{c|lcl|lcl|l}
$L$ & model layers           &  & $d_l$ & dense layers                     &  & $t$   & tensor parallel size   \\
$h_{d}$ & hidden dim             &  & $k_a$ & kv head number                   &  & $p$   & pipeline parallel size \\
$s$ & sequence length        &  & $e_n$ & intermediate size in moe layer   &  & $c$   & context parallel size  \\
$h$ & hidden size            &  & $g_d$ & intermediate size in dense layer &  & $e$   & expert parallel size   \\
$a$ & head number            &  & $g_e$ & intermediate size in moe layer   &  & $d$   & data parallel size     \\
$V$ & vocabulary size        &  & $t_k$ & topk                             &  & $b$   & micro batch size       \\
$l$ & model layers per stage &  & $v$  & pipeline stages per GPU          &  & $g_{bs}$ & global batch size     
\end{tabular}}
\end{table}

Although the methods mentioned above can currently effectively alleviate the phenomenon of expert load imbalance, they cannot solve the OOM problem caused by extreme distribution during training in large EP scenarios on GPUs with small memory capacity. Table~\ref{notationlist} lists symbols to be used in this paper.
\section{Theoretical Memory Cost Model}
\begin{figure}[t]
\centering
    \includegraphics[scale=0.5]{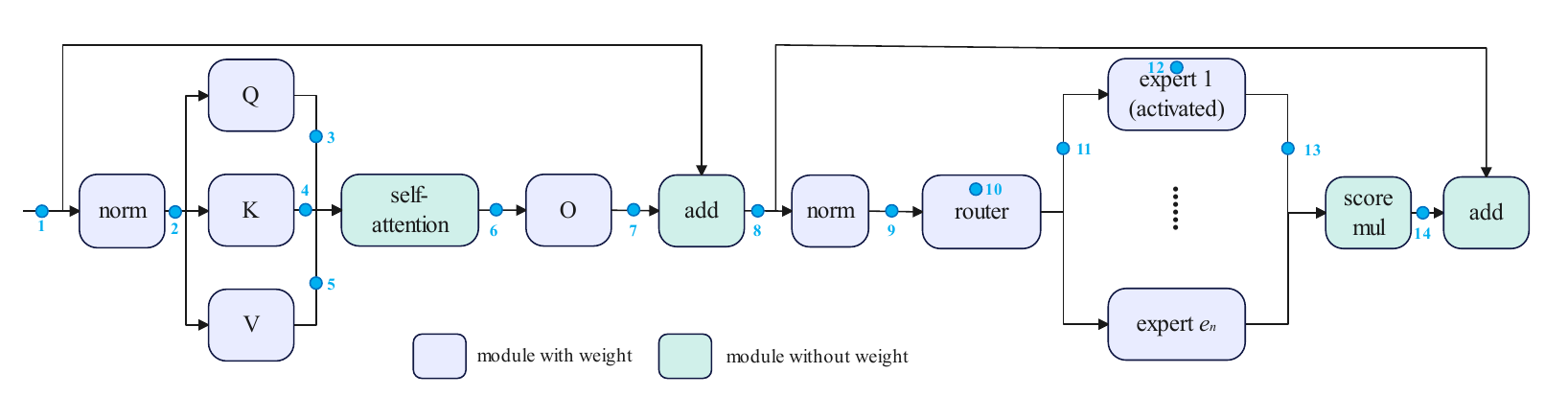}
    \caption{General Architecture of MoE Models. The blue dots denote the stored activation.}
    \label{moe-arc}
\end{figure}
\begin{table}[t]
\caption{Activation of MoE}
\centering
\label{moe-activation}
\renewcommand{\arraystretch}{1.1}
\scalebox{0.8}{
\begin{tabular}{ccc}
\hline
module                            & input ID & stored activation (Bytes)                                 \\ \hline
norm                              & 1        & $D_tbsh/(tc)$                                     \\
q, k, v                           & 2        & $D_tbsh/(tc)$                                     \\
\multirow{3}{*}{attention}        & 3        & $D_tbsah_d/(tc)$                                 \\
                                  & 4        & $D_tbsk_ah_d/(tc)$                                \\
                                  & 5        & $D_tbsk_ah_d/(tc)$                                  \\
o                                 & 6        & $D_tbsh/(tc)$                                     \\
add                               & 7        & -                                                 \\
norm                              & 8        & $D_tbsh/(tc)$                                    \\
\multirow{2}{*}{router}           & 9        & $D_tbsh/(tc)$                                     \\
                                  & 10       & $D_tbse_n/(tc)$                                    \\
\multirow{2}{*}{activated expert} & 11       & $D_tbs'h/(tc)$                                    \\
                                  & 12       & $2D_tbs'g_e/(tc)$                                 \\
score mul                         & 13       & $D_tbs'h/(tc)$                                    \\
add                               & 14       & -                                                 \\ \hline
Total                             & -        & $\frac{1}{tc}[D_tbs(5h + ah_d + 2k_ah_d + e_n)$ \\
                                  &          & $+\; D_tbs'(2h+2g_e)]$ \\ \hline
\end{tabular}}
\end{table}
This section introduces the theoretical memory cost model for MoE to elucidate the memory usage trends during MoE training. 
Generally, the GPU memory occupied during MoE training can be divided into two components~\cite{zhao2025mofa}:

\textbf{Static Memory:} The static GPU memory occupied by the MoE model weights. This can be modeled as follows:
\begin{align}
\label{staticmem}
M^{\mathrm{sta}} = \underbrace{D_t^{\mathrm{para}} v l \sum_i^{m_n} S_i^{\mathrm{para}}}_{\mathrm{Parameters}} + \underbrace{D_t^{\mathrm{grad}} v l \sum_i^{m_n} S_i^{\mathrm{para}}}_{\mathrm{Gradients}} + \underbrace{4 D_t^{\mathrm{opt}} v l \sum_i^{m_n} S_i^{\mathrm{para}}}_{\mathrm{Optimizer\ States}}
\end{align}
where $S_i^{\mathrm{para}}$ denotes the weight size of the $i$-th module, and $m_n$ represents the number of modules with weights. Fig.~\ref{moe-arc} illustrates the general architecture of an MoE model. Blue blocks indicate modules with weights, while green blocks denote weightless modules.

\textbf{Activated Memory:} The activated GPU memory occupied by intermediate data generated during MoE model training, primarily originating from cascaded transformer layers. The blue dots in Fig.~\ref{moe-arc} represent the activation values stored for one MoE transformer layer. The storage size is detailed in Table~\ref{actmem}. Based on this, we model the peak activated memory usage during MoE model training as follows:
\begin{equation}
\label{actmem}
\begin{split}
M^{\mathrm{act}} &= \frac{m_g}{tc}D_tb\Biggl( s(5h + ah_d + 2k_ah_d+e_n) + s'(2h+2g_e)\Biggr)
\end{split}
\end{equation}
where $m_g$ denotes the stored number of activation memory, $D_t$ denotes the data precision. Generally, $m_g=\left(v p + p - 2 r_{pp} - 1 \right)$, and when full recomputation is employed, then $m_g=1$.

\begin{figure}[t]
\centering
    \includegraphics[scale=0.38]{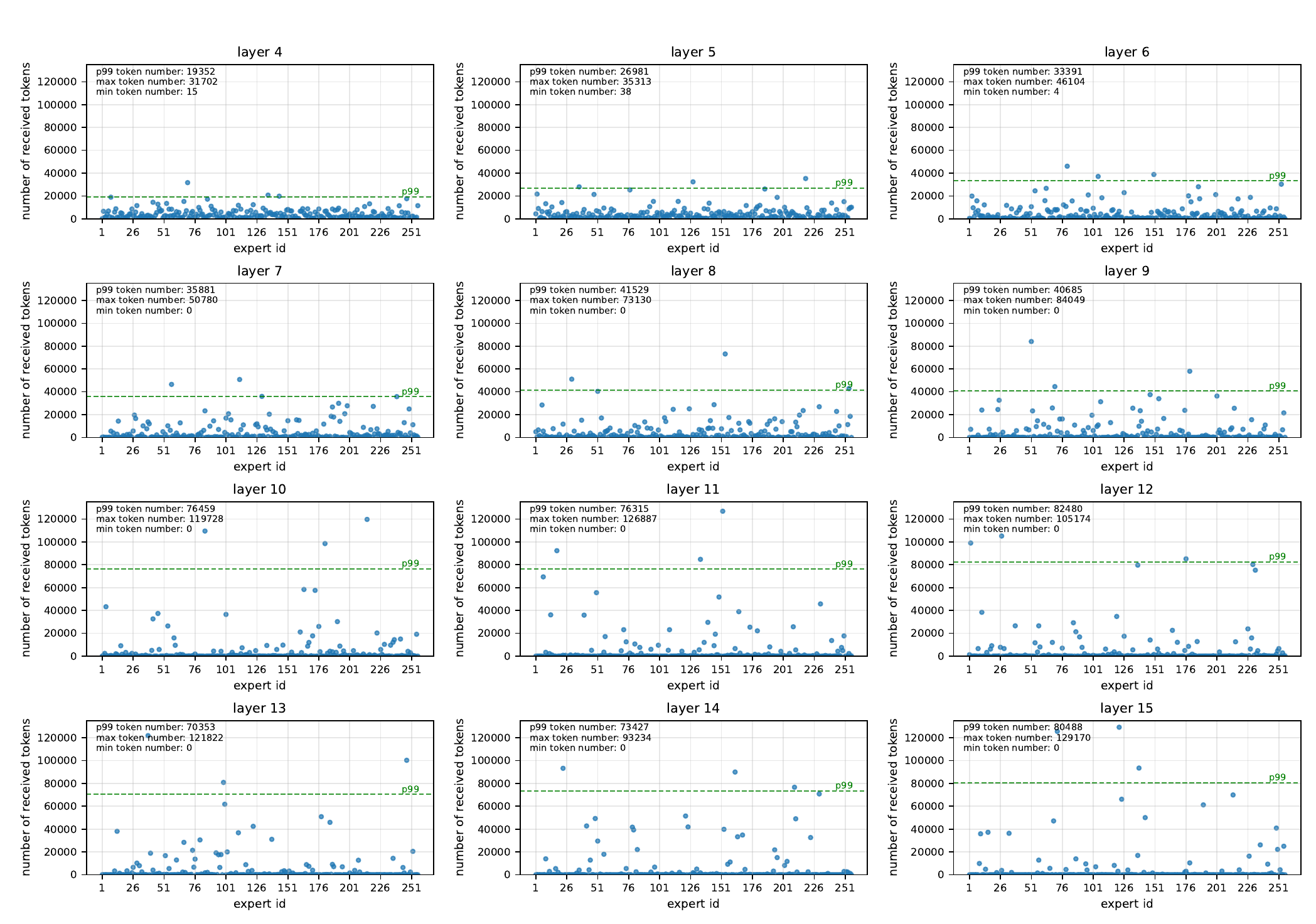}
    \caption{The number of received tokens per MoE layer. Take the 7-th iteration for an example.}
    \label{dispatch-analysis}
\end{figure}

Through equation~\eqref{actmem}, it can be observed that the peak activated memory is related to the sequence length $s$ and the number of tokens received by the MoE layer $s'$. During the early stages of MoE training when token distribution is uneven, there is a probability that a large number of tokens are routed to a single GPU, i.e., $s'$ approaches $es$. In large EP scenarios, this leads to extremely high activation memory usage. Even with full recomputation enabled, it can cause OOM errors, preventing normal training. Fig.~\ref{dispatch-analysis} displays the token distribution in the layer-reduced DeepSeek model used in our experiments. It shows that as the layer depth increases, the token distribution becomes increasingly uneven. In the latter layers, significant outliers appear, with the maximum number of received tokens approaching the theoretical peak and the minimum being zero. This phenomenon indicates that most activations are concentrated on a single GPU, leading to OOM.

To ensure normal model training, the following condition must be satisfied:
\begin{align}
\label{memlimit}
    M^{\mathrm{sta}}+M^{\mathrm{act}} \leqslant \alpha M^\mathrm{GPU}
\end{align}
where $M^\mathrm{GPU}$ denotes the GPU's specified memory capacity, and $\alpha$ is the available memory ratio for the model.
\section{Design of MemFine}
This section introduces MemFine from three aspects: the Fine-grained Chunk Distribution Algorithm and Memory-Aware Chunk Tuning.

\subsection{Fine-grained Chunk Distribution Algorithm (FCDA)}
\begin{figure}[t]
\centering
    \includegraphics[scale=0.38]{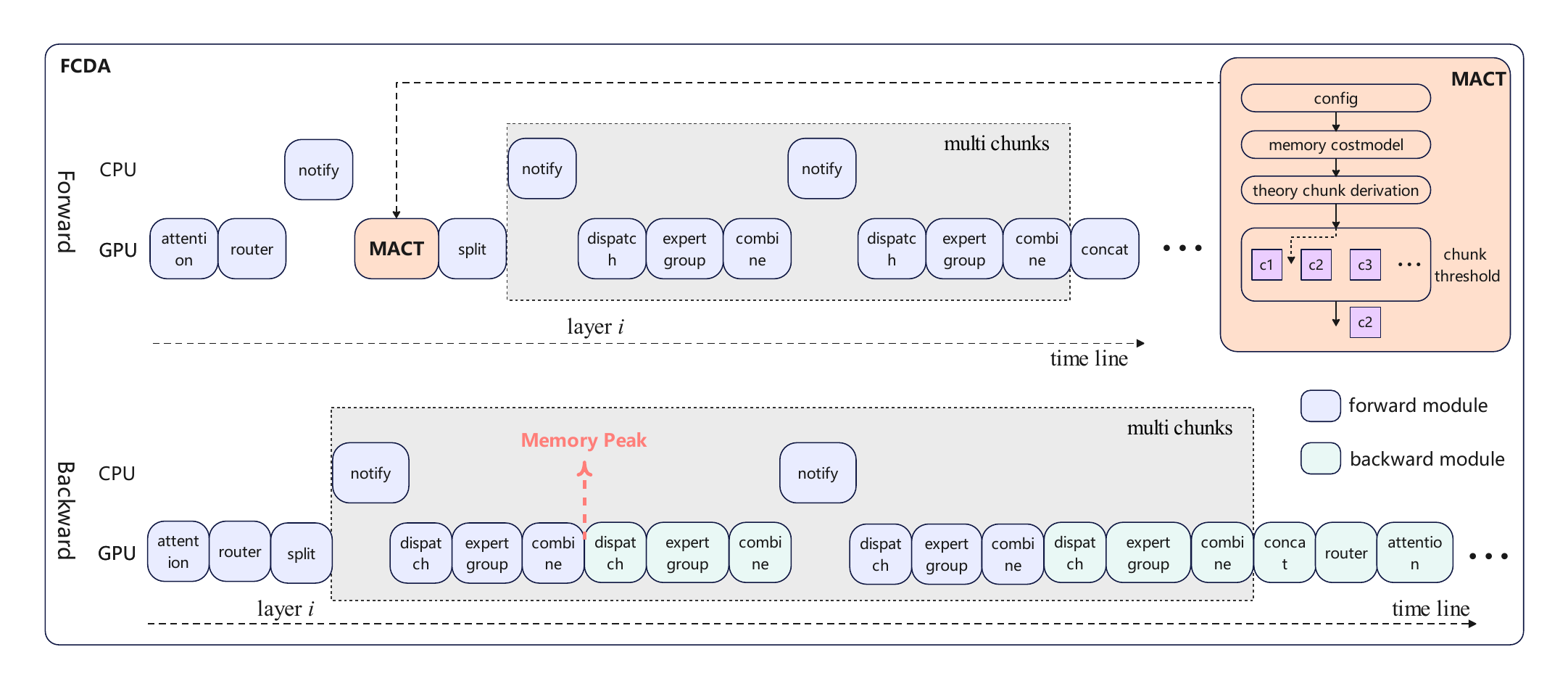}
    \caption{The workflow of MimFine.}
    \label{FCDA}
\end{figure}

In standard MoE model training, the forward process of dispatch-computation-combine and the backward process are expressed as follows:
\begin{align}
\label{fwd}
    Y = F_w(X)=\mathrm{combine}(\mathrm{expert}(\mathrm{dispatch}(X)))
\end{align}
\begin{align}
\label{bwd}
    X_{\mathrm{grad}} = B_w(Y_{\mathrm{grad}},F_w(X))
\end{align}
where $F_w$ and $B_w$ denote the forward propagation and the backward propagation, respectively. Under this paradigm, MoE models may still experience peak memory usage from a single layer's activations. In extreme dispatch scenarios, this can lead to OOM caused by $F_w(X)$, as discussed in Section 3. To address this issue, we propose the FCDA. The FCDA consists of two designs:

\textbf{Forward Propagation:} As shown in Fig.~\ref{FCDA}, we divide tokens into multiple token chunks. Each chunk is processed sequentially through dispatch-computation-combine operations. The final output is obtained by concatenating the results from all chunks. This process is formulated as:
\begin{align}
\label{fcda-fwd}
    Y = F_w(X)=\mathrm{concat}(F_w(X_1),F_w(X_2),...,F_w(X_c))
\end{align}
\textbf{Backward Propagation:} We redesign the recomputation scheduling mechanism based on a chunk-level recomputation and backward propagation. As shown in Fig.~\ref{FCDA}, the process is formulated as:
\begin{align}
\label{fcda-bwd}
    X_{\mathrm{grad}} = \mathrm{concat}(B_w(Y_{\mathrm{grad}},F_w(X_1)),B_w(Y_{\mathrm{grad}},F_w(X_2)),...,B_w(Y_{\mathrm{grad}},F_w(X_c)))
\end{align}

Through this approach, we can reduce the peak memory usage of MoE layers to the maximum activation value among chunks. The memory reduction amounts to $F_w(X)-\max(F_w(X_1),F_w(X_2)),...,F_w(X_c)))$.

\begin{table}[t]
\caption{Model configuration}
\centering
\label{model}H
\renewcommand{\arraystretch}{1.2}
\begin{tabular}{c|cccccccccc}
\hline
\diagbox{model}{config} & $L$ & $s$  & $h$  & $a$ & $g_d$ & $g_e$ & $t_k$ & $V$    & $r$  & $d_l$ \\ \hline
model I                                      & 16  & 4096 & 7168 & 128 & 18432 & 2048  & 8     & 129280 & 1536 & 3     \\
model II                                     & 8   & 4096 & 7168 & 128 & 18432 & 2048  & 8     & 129280 & 1536 & 3     \\ \hline
\end{tabular}
\end{table}
\begin{table}[t]
\caption{Memory comparison}
\label{mem-campare}
\centering
\renewcommand{\arraystretch}{1.2}
\begin{tabular}{cccccc}
\hline
model               & \multicolumn{1}{c}{method}               & static mem. (GB) & active mem. (GB) & all mem. (GB) & training \\ \hline
\multirow{3}{*}{I}  & 1  & 43.0  & 22.9 & 65.9 & $\times$            \\
                    & 2  & 43.0  & 3.7 & 46.7 & $\checkmark$             \\
                    & 3  & 43.0  & 11.9 & 54.9 & $\checkmark$             \\ \hline
\multirow{3}{*}{II} & 1  & 39.5  & 22.9 & 62.4 & $\checkmark$ \\
                    & 2  & 39.5  & 3.7 & 43.2 & $\checkmark$ \\
                    & 3  & 39.5  & 11.9 & 51.4 & $\checkmark$ \\ \hline
\end{tabular}
\end{table}

\subsection{Memory-Aware Chunk Tuning (MACT)}
Given that different Pipeline Parallelism (PP) stages may be deployed, we observe varying memory pressure across PP stages, which consequently affects chunk size. Fixed chunk sizes pose potential OOM risks and impact training performance. To address this challenge, we further propose Memory-Aware Chunk Tuning (MACT). The design concept of MACT is illustrated in the upper-right portion of Fig.~\ref{FCDA}.

First, before training, the MACT system models the training memory usage based on the model configuration. It then calculates the theoretical maximum $s'$ for different PP stages as follows:
\begin{align}
\label{smax}
    s'_{\mathrm{max}}=\frac{\alpha M^\mathrm{GPU}-M^{\mathrm{sta}}-\frac{m_g}{tc}D_tbs(5h + ah_d + 2k_ah_d+e_n)}{\frac{m_g}{tc}D_tb(2h+2g_e)}
\end{align}

Based on the result of the first notification, we can get the number of received tokens per GPU, denoted as $s''$, and derive the theoretically optimal chunk value:
\begin{align}
\label{theory-c}
    c=\left\lceil s''/s'_{\mathrm{max}} \right\rceil
\end{align}

Considering that introducing \eqref{smax} and \eqref{theory-c} would increase the computational cost, we use a threshold method for the selection of chunk size. In particular, MACT firstly categorizes chunk sizes into several bins, and then select the large bin that is closest to $c$ as the chunk size setting. During training, MACT dynamically adjusts the chunk size to balance memory usage and performance.
\section{Experiment and Analysis}
\subsection{Experimental Setup}
Our experiments were conducted on 32 GPUs with memory 64GB per GPU. Two reduced-layer models based on DeepSeek-V3 were selected, as detailed in Table~\ref{model}. Specifically, we applied a parallelism strategy of $t=1$, $p=4$, $e=32$, $d=1$, $c=1$, $l=L$, $v=1$, $b$=1, and $g_{bs}=960$. BF16 data precision is used in pretraining, i.e., $D_t=2$.
The distributed framework employed was Megatron-LM, and the software environment consisted of Python 3.8 and PyTorch 2.1.0.

\begin{figure}[t]
    \includegraphics[scale=0.46]{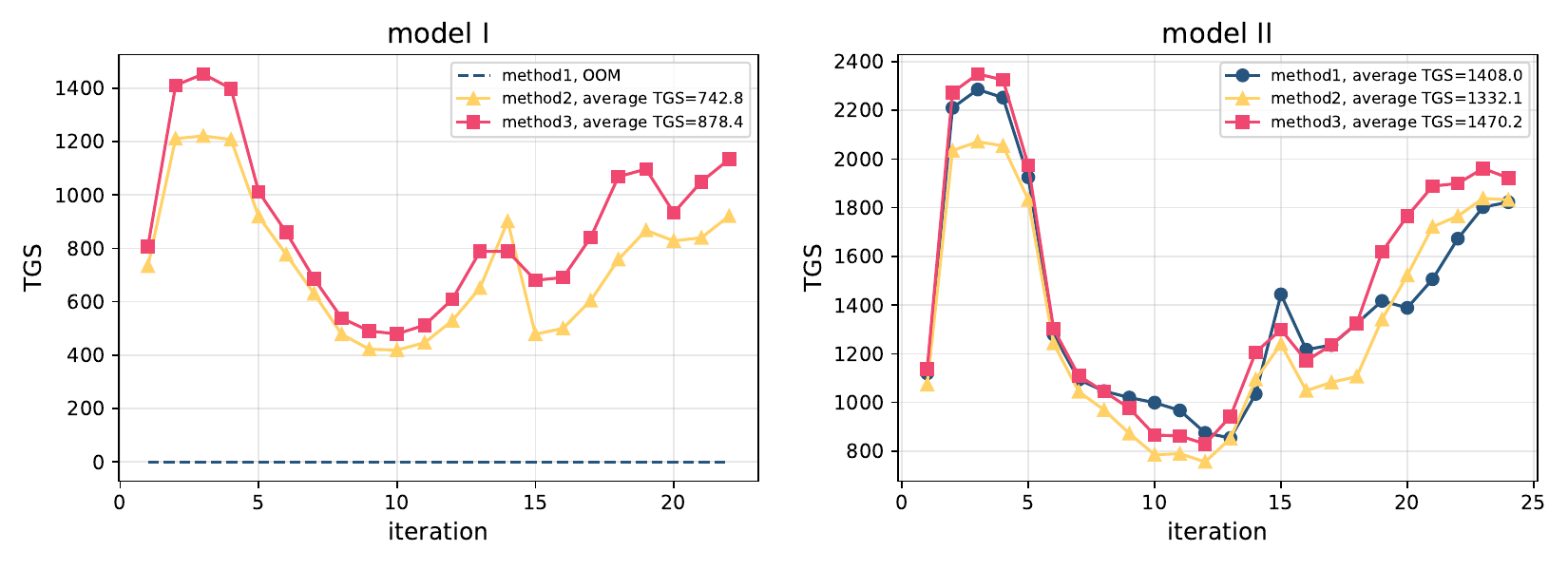}
    \caption{Throughput comparison of three methods.}
    \label{throughput}
\end{figure}

\begin{figure}[t]
\centering
    \includegraphics[scale=0.41]{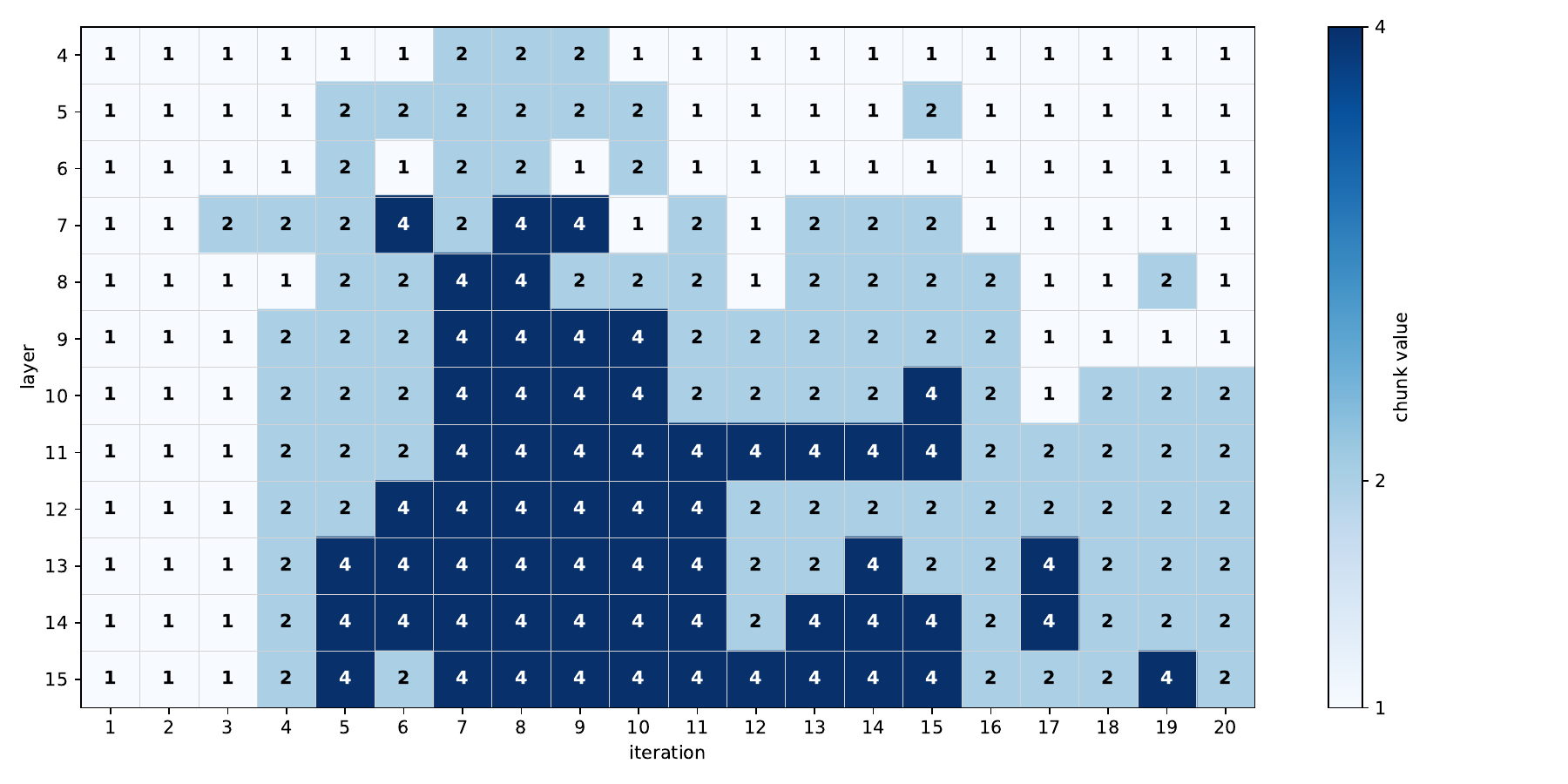}
    \caption{Trend of chunk values during training of Model I with Method 3.}
    \label{chunk-change}
\end{figure}

First, we compared the memory consumption of several optimization methods, including:
\begin{enumerate}[label=\textbullet]
    \item Method 1: Training without chunk splitting and applying full recomputation to save memory, which is widely used in Megatron-LM~\cite{shoeybi2019megatron};
    \item Method 2: MemFine with fixed chunk threshold, e.g., $c_k=8$.
    \item Method 3: MemFine with MACT, where chunk threshold is set to [1,2,4,8].
\end{enumerate}

The memory consumption results are shown in the Table~\ref{mem-campare}. For Model I, Method 1 generates extreme activation memory usage, which leads to OOM, preventing normal continuous training. In comparison, MemFine with fixed $c_k=8$ effectively reduces activation memory by $83.84\%$. Under the MACT algorithm, MemFine derives an optimal $c_k=2$, which also reduces activation memory by $48.03\%$. Similar conclusions can be drawn for Model II.

On the other hand, we compared the training performance of the three methods on both models, as shown in Fig.~\ref{throughput}. Overall, the performance on both models initially improves, then decreases, and gradually improves again, which correlates with token distribution patterns in the early training stages. Specifically, for Model I, Method 3 achieves the best performance, with an average tokens per gpu per second (TGS) improvement of $18.26\%$ compared to Method 2, while Method 1 cannot train normally due to OOM. In particular, TGS is calculated as follows
\begin{align}
    TGS = \frac{g_{bs}s}{TN}
\end{align}
where $T$ denotes as the time of one iteration and $N$ denotes as the number of GPU. For Model II, Method 3 improves average TGS by $4.42\%$ compared to Method 1, while Method 2 shows a $5.40\%$ degradation in average TGS compared to Method 1. The mitigation of OOM issues and performance improvements fully demonstrate the effectiveness of the design of the MemFine with MACT.

Finally, using Model I as an example, we analyzed the effectiveness of MACT, as shown in the Fig.~\ref{chunk-change}. It can be observed that as training iterations increase, the dynamically selected chunks in Model II first increase and then decrease. Larger chunks are concentrated in layers 7-15 during iterations 5-15. Furthermore, later layers exhibit more large chunk values, indicating that the model is in a relatively chaotic state during these iterations, with weaker feature learning capability in later layers leading to extremely uneven token distribution. After approximately 10 iterations, the distribution begins to stabilize, suggesting the model is gradually learning expert features from the chaotic state.
\section{Conclusion}
To addresses the critical memory bottleneck in large-scale MoE training caused by dynamic token routing and severe load imbalance, this paper proposed MemFine, a memory-aware fine-grained scheduling framework that effectively eliminates memory overflow on capacity-constrained GPUs without compromising the dynamic routing mechanism essential for model accuracy. By decomposing token distribution and expert computation into manageable chunks and employing a dynamically optimized chunked recomputation strategy, MemFine significantly reduces activation memory usage while maintaining high training throughput. Experimental results demonstrate that our method reduces activation memory by $48.03\%$ and improves throughput by $4.42\%$ compared to full recomputation baselines. MemFine thus provides an efficient and practical solution for enabling stable training of ultra-large-scale MoE models on memory-limited hardware, opening new possibilities for scaling AI models efficiently. Future work will explore the application of MemFine to broader distributed training scenarios and more diverse model architectures.
\section{Authors}

\textbf{ZTE AIH Team}: 

Lu Zhao, Rong Shi, Shaoqing Zhang\footnote[2]{Corresponding Authors. \quad Email: zhang.shaoqing1@zte.com.cn,\quad sun.hongfeng@zte.com.cn}, Yueqiang Chen, Baoguo He, Hongfeng Sun\footnotemark[2], Ziqing Yin, Shangchao Su, Zhiyan Cui, Liang Dong, Xiyuan Li, Lingbin Wang, Jianwei He, Jiesong Ma, Weikang Huang, Jianglei Tong, Dongdong Gao, Jian Zhang, Hong Tian.

\textbf{INESA-ZTE Joint Laboratory}: 

Hui Shen, Zongtai Luo, Zhaoqun Sun, Hongxing Niu, Yue Sun.
\section*{Acknowledgement}
We wish to express our sincere gratitude to the INESA-ZTE Joint Laboratory, which kindly provided the essential test environment and valuable technical assistance throughout the research and manuscript preparation process.

{\small
\bibliographystyle{unsrt}%
\bibliography{nips21}
}

\end{document}